\documentstyle[a4,sprocl,epsfig]{article}
\def\Journal#1#2#3#4{{#1} {\bf #2}, #3 (#4)}

\def\NIMA{{\em Nucl. Instrum. Methods} A}

\def\PLB{{\em Phys. Lett.}  B}
\def\PRL{\em Phys. Rev. Lett.}
\def\PRD{{\em Phys. Rev.} D}
\def\ZPC{{\em Z. Phys.} C}
\def\CPC{\em Comp. Phys. Comm.}

\def\be{\begin{equation}}
\def\ee{\end{equation}}
\def\bea{\begin{eqnarray}}
\def\eea{\end{eqnarray}}

\begin{document}
\begin{titlepage}

\begin{center}
UNIVERSITY OF HELSINKI \hspace{2.0cm} REPORT SERIES IN PHYSICS
\end{center}

\begin{center}
\vspace{0.8 cm}


HU-P-264

\vspace{3.0 cm}
\Large
{\bf Measuring Higgs Branching Ratios} \\ 
\vspace{0.25cm}
{\bf and telling the SM from a MSSM Higgs Boson}\\
\vspace{0.25cm}
{\bf at the $e^+e^-$ Linear Collider }
\vspace{1.5 cm}
\large

Marco Battaglia
                  
\vspace{0.5 cm}
\normalsize
Department of Physics, High Energy Physics Division\\
University of Helsinki\\
P.O. Box 9, FIN-00014 University of Helsinki, Finland \\

\vfill

To appear in the Proceedings of the\\  
{\sl International Workshop on Linear Colliders LCWS99},\\
Sitges (Spain),  April 28 - May 5, 1999

\end{center}
\end{titlepage}

\clearpage
\pagestyle{empty}
\normalsize

\mbox{$ \left. \right. $}

\vfill

\begin{center}
ISBN 951-45-8195-4\\
ISSN 0355-5801 \\
\vspace*{0.5cm}
Helsinki 1999 \\
\end{center}

\clearpage 
\newpage
\pagestyle{plain}
\setcounter{page}{1}

\title{MEASURING HIGGS BRANCHING RATIOS\\
AND TELLING THE SM FROM A MSSM HIGGS BOSON\\
AT THE $e^+e^-$ LINEAR COLLIDER}

\author{M. Battaglia}

\address{Department of Physics, High Energy Physics Division
\\ University of Helsinki (Finland)}

\maketitle\abstracts{An accurate determination of the Higgs decay 
branching ratios to 
$b \bar b$, $c \bar c$, $g g$, $W W^*$ and $\tau^+ \tau^-$ pairs is important 
for the study of the Higgs couplings and for determining its 
SM or MSSM nature in the mass range 100~GeV/c$^2 < M_H < $
140~GeV/c$^2$. This measurement also represents an important benchmark for the
optimisation of the detector design. The accuracy on the determination
of the Higgs decay branching ratio to fermions and $W W^*$ pairs has been 
studied using the simulation of the detector designed for the TESLA $e^+e^-$ 
linear collider at $\sqrt{s}$ of 350~GeV and 500~GeV. The results are 
discussed in terms of the ability of discriminating between the SM and a MSSM 
neutral Higgs boson and of the predictivity on the $A^0$ mass in the MSSM 
scenario.}

\section{Introduction}
The investigation of the production and decay properties of the Higgs boson 
and the determination of its nature are an important part of the physics 
programme at the $e^+e^-$ linear collider.
Furthermore this study sets stringent requirements on the response
of the experimental apparatus and thus represents an ideal benchmark for the 
optimisation of its design.

At the linear collider, detection of the SM or MSSM-like neutral Higgs boson
will be straightforward and the anticipated large statistics and accurate
detector response will enable detailed tests of its production and decay 
characteristics~\cite{janot}.

This study is based on the simulated response of the detector designed for
the {\sc Tesla} CDR~\cite{cdr}. The track impact parameter resolution has been 
varied to reflect the updates in the Vertex Tracker design~\cite{vt}.
Signal $e^+e^- \rightarrow Z^0H^0$, $H^0 \nu \bar \nu$ and background events
have been simulated using the {\sc Pythia~5.02} and {\sc Jetset~7.405}
generators~\cite{jetset} tuned on {\sc Lep} data for the 
electroweak and QCD variables and on both {\sc Lep} and {\sc Cleo} data for
heavy flavour decays. Events have been generated at $\sqrt{s}$ of
350~GeV and 500~GeV, corresponding to an integrated luminosity of 
$\int L =$ 100~fb$^{-1}$. The centre-of-mass energy included the smearing 
effect due to beamstrahlung~\cite{daniel}. The results have been scaled to the
integrated luminosities of 500~fb$^{-1}$ corresponding to 1 to 2~years 
(1 year = $10^7$ s) of data taking at the {\sc Tesla} design luminosity of 
$3 \times 10^{34}$~cm$^{-2}$ s$^{-1}$.
The detector response to charged particle tracks has been studied using a full 
{\sc Geant~3.21} simulation and a track fit based on the Kalman filter 
algorithm.
The resulting momentum and impact parameter resolutions have been used as 
inputs to a parametric smearing simulation program~\cite{simdet}.
This program has been also used to model the calorimeter response. 
Events have been reconstructed using a dedicated event reconstruction and 
analysis program. Finally a jet flavour tagging algorithm has been applied in 
order to identify the hadronic final states. The reconstruction, analysis and 
tagging algorithms are based on programs developed for the study of the 
{\sc Lep-2} data with the {\sc Delphi} detector.

\section{Determination of Higgs Branching Ratios}

The accuracy on the Higgs decay branching ratio measurements at the linear 
collider has been already the subject of several 
studies~\cite{hildreth}$^{\!-\,}$\cite{desy123}.
With improved designs of the Vertex Tracker, more advanced jet flavour tagging
techniques, profiting of the experience gained at {\sc Lep}
and {\sc Slc}, and the large statistics available at the {\sc Tesla} collider
these studies move in the domain of precision measurements.

\subsection{Event Reconstruction}

This analysis considered neutral Higgs bosons, produced by either the 
$e^+e^- \rightarrow H^0 Z^0$ or the $e^+e^- \rightarrow H^0 \nu \bar \nu$
processes, in the $(JJ)_H (jj)_Z$, $(JJ)_H (\ell \ell)_Z$ and 
$(JJ)_H \nu \bar \nu$ final states, where $(JJ)_H$ stands both for the two jet
hadronic final states and for $\tau^+ \tau^-$ and $W W^*$ as discussed below.

After the event selection, hadronic jets have been reconstructed using the 
{\sc Luclus} algorithm~\cite{jetset}. Events have been forced to either two or
four jets and the jet energies have been rescaled using a 4-C fit by imposing 
energy and momentum conservation at the nominal $\sqrt{s}$. 
Further selections have been applied on the event topology
for the three final states under study. Finally the compatibility of each 
reconstructed event with the $H^0 Z^0$, $H^0 \nu \bar \nu$ hypotheses
has been tested by constructing a $\chi_M^2$ variable based on the
reconstructed masses. 
The di-jet Higgs mass resolution has been found to be 
$\sigma_{JJ} \simeq 2.7$~GeV/$c^2$ at 120~GeV/$c^2$. 
Candidate Higgs decays have been selected by a cut on this  
$\chi_M^2$ variable. The efficiency of this selection has been estimated to 
be 25\% with 76\% purity, corresponding to a signal to background ratio 
$S/\sqrt{B} \simeq 100$, at both $\sqrt{s}$ = 350~GeV and 500~GeV. 

The Higgstrahlung cross-section $\sigma_{ZH}$ can be obtained by the study 
of the di-lepton recoil mass, $M_{\ell \ell}$, from $Z^0 \rightarrow 
\ell^+ \ell^-$. This is necessary for the extraction of the absolute Higgs 
decay branching ratios by flavour tagging of the Higgs decay final states. 
In addition, a determination of the Higgs production 
cross-section independent on its decay modes is important in the study
of possible invisible or exotic Higgs decay channels such as decays to SUSY 
A dedicated study has shown that an accuracy of about 2\% on $\sigma_{ZH}$ can 
be obtained for both $M_H$ = 120~GeV/$c^2$ and $M_H$ = 140~GeV/$c^2$ with 
an integrated luminosity of 500~fb$^{-1}$. 

\subsection{$H^0 \rightarrow b \bar b$, $c \bar c$, $g g$}

Higgs hadronic final states have been selected by imposing hadronic jet
quality cuts.
The separation of the $b \bar b$, $c \bar c$, $g g$ hadronic final states is
based on the jet flavour tagging response. 
Due to their relatively long lifetime, cascade $b \rightarrow c \rightarrow s$
decay chain and large invariant mass of beauty hadrons, the decay topology 
with two secondary vertices and the secondary particle invariant mass are 
very distinctive features of $b$~jets. 
At the linear collider, due to the large boost, the $b$~hadrons decay on 
average about 0.8~cm away from the beam collision, compared to 
0.3~cm at {\sc Lep} and at the {\sc Slc}. Furthermore the 
performances of the new generation of Vertex Trackers presently under study
for the linear collider will allow an efficient reconstruction of secondary
charged particle and the determination of their point of production. Jet 
flavour tagging techniques, similar to those already successfully applied 
at {\sc Lep}~\cite{taglep} and at the {\sc Slc}~\cite{tagslc}, will ensure 
effective performances due to the favourable kinematics and detector response.
The tagging algorithm adopted in this analysis combines kinematical
and topological variables in order to identify $b \bar b$, $c \bar c$ and 
light quark di-jets.
\begin{figure}[hb!]
\begin{center}
\epsfig{file=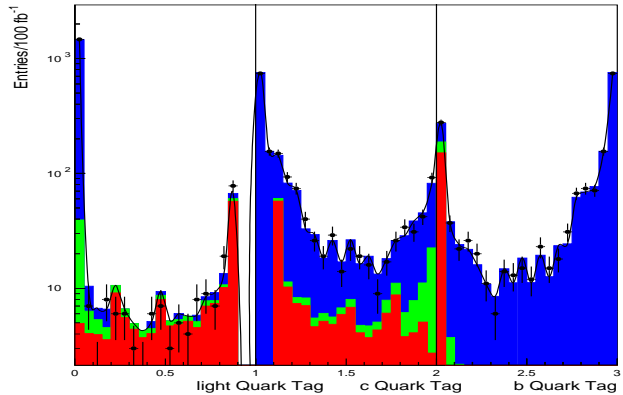,bbllx=5,bblly=150,
bburx=560,bbury=660,width=9.0cm,height=5.5cm}
\end{center}
\caption{The distribution of the $gg$, $cc$ and $bb$ di-jet flavour tag 
response for Higgs decay candidates. The points with error bars show the 
expected data distribution assuming SM couplings for 100~fb$^{-1}$ and the 
shaded histograms the response for light quark (medium grey), $c \bar c$ 
(light grey) and $b \bar b$ (dark grey) di-jets normalised to the fitted
fractions. \label{fig:fit}}
\end{figure}

For each hadronic jet an inclusive secondary vertex search has been performed
using charged particles tracks reconstructed in the Vertex Tracker. 
Neutral particles with large probability of originating from a heavy hadron 
decay have been also associated to this secondary vertex. The identification 
of the jet flavour has been based on the following variables:
i) probability that all the reconstructed charged particles of the jet
originated at the event primary vertex, ii) number of charged particles 
with an impact parameter larger than twice the measurement accuracy, 
iii) invariant mass and energy of the secondary particles, iv)
transverse momentum of identified leptons in the jets and v) probability 
that all the secondary particles originated at a common, displaced vertex.
The likelihood function has been defined as follows: for each of the $N$ 
discriminating variables, the fractions  $F_{i}^{bb}(x_i)$, $F_{i}^{cc}(x_i)$
and $F_{i}^{gg}(x_i)$ of $b$, $c$ and $g$ di-jets corresponding 
to a given value $x_i$ of the $i^{th}$ variable, have been extracted
from samples of Higgs $b \bar b$, $c \bar c$ and $gg$ decays with equal 
populations.
The $bb$, $cc$ and $gg$ likelihood variables have been computed as the 
normalised product of these individual fractions as: 
$\prod_{i=1,N} F_{i}^{QQ}(x_i)/\sum_{qq=bb,cc,gg} \prod_{i=1,N} 
F_{i}^{qq}(x_i)$. 
Only decays with at least one jet with jet flavour tagging response have been 
considered in this analysis. 
The fractions of $b \bar b$, $c \bar c$ and $g g$ Higgs final states has been 
extracted by a binned maximum likelihood fit to these di-jet flavour tagging 
probabilities for the Higgs decay candidates (see Figure~\ref{fig:fit}). 
The background 
has been estimated from simulation over a wide interval around the Higgs mass 
peak and subtracted. It will be possible to study the flavour composition of 
this background directly on the real data by using the side-bands of the Higgs
mass peak, thus reducing the possible systematics error from the simulation
modelling. The fitted fractions agreed well with the input values and the 
performances at $\sqrt{s}$ = 350~GeV and 500~GeV have been found to be 
equivalent.
\begin{table}[hb!]
\begin{center}
\caption{Accuracy in the determination of the hadronic Higgs 
decay branching ratios for $M_H$ = 120~GeV/c$^2$ and $\sqrt{s}$ = 350~GeV
\label{table:qq}}
\vspace{0.4cm}
\begin{tabular}{|c|c|}
\hline
Channel & $\delta (\frac{\sigma_{ZH} \times BR(H \rightarrow X)}
{\sigma_{ZH} \times BR(H \rightarrow {\mathrm hadrons})}) / BR$\\
\mbox{ } &~~~~~CDR Vtx.~~$|$~~Improved Vtx. \\
\hline \hline
$H^0 / h^0 \rightarrow b \bar b$    & $\pm$ 0.011~~$|$~~$\pm$ 0.008 \\
$H^0 / h^0 \rightarrow c \bar c$    & $\pm$ 0.134~~$|$~~$\pm$ 0.080 \\
$H^0 / h^0 \rightarrow g g$         & $\pm$ 0.050~~$|$~~$\pm$ 0.050 \\
\hline
\end{tabular}
\end{center}
\end{table}

\subsection{$H^0 \rightarrow \tau^+ \tau^-$}

The selection of $H^0 \rightarrow \tau^+ \tau^-$ decays started from events
not fulfilling the Higgs hadronic final state selection. 
Events have been clusterised in either two or four jets with isolated tracks 
being considered as single particle jets. 
\begin{table}[hb!]
\begin{center}
\caption{Performances of the $H \rightarrow \tau \tau$ 
analysis for $M_H$ = 120~GeV/c$^2$ and $\sqrt{s}$ = 350~GeV \label{table:tau}}
\vspace{0.4cm}
\begin{tabular}{|c|c|c|c|}
\hline
Signal $H \rightarrow \tau \tau$ & Background & Eff. &
$\delta BR/BR$ \\
/100 fb$^{-1}$ & /100 fb$^{-1}$ & \mbox{ } & /500 fb$^{-1}$\\
\hline \hline
165 &  280 & 19\% & $\pm$ 0.057 \\ \hline
\end{tabular}
\end{center}
\end{table}
Similarly to the case of the jet flavour tagging of hadronic final states, a 
global $\tau \tau$ likelihood has been defined by using the response of the 
following discriminating variables: 
i) number of jets (including isolated particles), ii) number of jets 
with more than 2 charged particles in the event, iii) event Thrust, iv) 
$E_{tot} / \sqrt{s}$, v) jet invariant mass, vi) number of charged particles 
per jet and vii) jet impact parameter probability. This $\tau \tau$ likelihood
peaked at one for $H^0 \rightarrow \tau^+ \tau^-$ decays and at zero for
background events. $H \rightarrow \tau \tau$ decays have been selected by a 
cut on this likelihood. The background from other Higgs 
decay channels has been found to be negligeable. Results are summarised in 
Table~\ref{table:tau}. 

\subsection{$H^0 \rightarrow W W^*$}

The determination of the $W W^*$ decay branching ratio has been performed 
using the $H^0 \rightarrow W W^* \rightarrow \ell \bar \nu q \bar q$
channel. This analysis of the $H^0 \rightarrow W W^*$ is presented in details 
in~\cite{ww}. The signal is characterised by the event topology and the
two-jet recoil mass distribution peaked at the Higgs mass. The results are
summarised in Table~\ref{table:w}.

\begin{table}[h!]
\begin{center}
\caption{Performances of the $H \rightarrow W W^*$ 
analysis for $M_H$ = 120~GeV/c$^2$ and $\sqrt{s}$ = 350~GeV \label{table:w}}
\vspace{0.4cm}
\begin{tabular}{|c|c|c|c|}
\hline
$H \rightarrow W W^*$ & Background & Eff. &
$\delta BR/BR$ \\
/100 fb$^{-1}$ & /100 fb$^{-1}$ & \mbox{ } & /500 fb$^{-1}$ \\
\hline \hline
101 & 30 & 5.3\% & $\pm$ 0.051 \\ \hline
\end{tabular}
\end{center}
\end{table}

This measurement, combined with the determination of the Higgstrahlung and 
$WW$ fusion cross-sections, can be used to extract the Higgs width with good
accuracy~\cite{ww}.
 
\section{Telling a SM from a MSSM Higgs Boson}

The accuracies on the determination of the decay branching ratio into 
$b \bar b$, $c \bar c$, $g g$, $\tau \tau$ and $W W^*$, obtained in this study
for the case of a $M_H$ = 120~GeV/c$^2$ Higgs boson, are summarised in 
Table~\ref{table:sum}.
\begin{table}[ht!]
\begin{center}
\caption{Accuracy in the determination of the Higgs decay
branching ratios \label{table:sum}}
\vspace{0.4cm}
\begin{tabular}{|c|c|}
\hline
Channel & $\delta (BR(H \rightarrow X) / BR$ \\
\mbox{ } &~~~~~CDR Vtx.~~$|$~~Improved Vtx. \\
\hline \hline
$H^0 / h^0 \rightarrow b \bar b$       & $\pm$ 0.024~~$|$~~$\pm$ 0.024 \\
$H^0 / h^0 \rightarrow c \bar c$       & $\pm$ 0.135~~$|$~~$\pm$ 0.083 \\
$H^0 / h^0 \rightarrow g g$            & $\pm$ 0.055~~$|$~~$\pm$ 0.055 \\
$H^0 / h^0 \rightarrow \tau^+ \tau^-$  & $\pm$ 0.060 \\
$H^0 / h^0 \rightarrow W W^*$          & $\pm$ 0.054 \\
\hline
\end{tabular}
\end{center}
\end{table}
Their implications on distinguishing the SM or MSSM nature of the Higgs boson
are discussed in this section. 

The sensitivity of the Higgs decay branching ratios in discriminating between 
the SM and MSSM Higgs hypotheses has been already studied in some 
details~\cite{smssm}. In MSSM the Higgs decay width to specific final states
has the following dependence:
$\Gamma^{MSSM}_{b \bar b} \propto \Gamma^{SM}_{b \bar b} 
\frac{\sin^2 \alpha}{\cos^2 \beta}$,
$\Gamma^{MSSM}_{c \bar c} \propto \Gamma^{SM}_{c \bar c} 
\frac{\cos^2 \alpha}{\sin^2 \beta}$ and
$\Gamma^{MSSM}_{WW} \simeq \Gamma_{WW}^{SM}$, with 
 $\tan \alpha = \frac{(M^2_Z + M^2_A) \sin \beta \cos \beta}
{M^2_h - (M^2_Z \cos^2 \beta + M^2_A \sin^2 \beta)}$ and
$\tan \beta = \frac{v_2}{v_1}$ being the ratio of vacuum expectation values. 
Therefore deviations in the ratio of branching ratios such as
{$\frac{BR(h \rightarrow W W^*)}{BR(h \rightarrow b \bar b)}$,
$\frac{BR(h \rightarrow c \bar c)}{BR(h \rightarrow b \bar b)}$ and
$\frac{BR(h \rightarrow g g)}{BR(h \rightarrow b \bar b)}$ 
from their SM expectations could reveal the MSSM nature of the Higgs boson
and provide indirect information on the mass of the CP odd $A^0$ Higgs boson.
The accuracy of the SM predictions has been studied assuming 
$m_b(M_{\Upsilon(1S)}/2)$ = (4.20 $\pm$ 0.11)~GeV/$c^2$ for the running
$\overline{\mathrm MS}$, $m_b - m_c$ = (3.40 $\pm$ 0.04)~GeV/$c^2$,
{$\alpha_s (m_Z)$ = 0.1164 $\pm$ 0.0025 and $m_{top}$ = 
(175 $\pm$ 0.2)~GeV/$c^2$ where the $b$ quark mass has been obtained
using a sum rule for the $\Upsilon$ spectroscopy data~\cite{mb}, the 
$m_b - m_c$ mass difference has been derived from  Heavy Quark Effective 
Theory~\cite{mc} while for the top pole mass, $m_{top}$, the uncertainty 
reflects a conservative estimate of the expected accuracy from the top 
threshold scan at the linear collider~\cite{top}. 
\begin{figure}[ht!]
\begin{center}
\epsfig{file=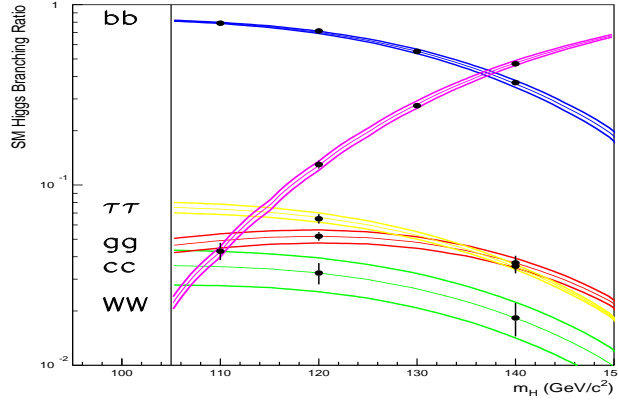,width=9.0cm,height=6.0cm}
\end{center}
\caption{The predicted SM Higgs decay branching ratios shown as the 68\% 
confidence level bands with overlayed the measured points using the results of
this study. \label{fig:sum}}
\end{figure}
Improvements on $m_b$ and $m_b - m_c$, possibly by a factor $\simeq 2$, can be 
envisaged after the study of the data on $B$-decays from the $B$-factories 
and the {\sc Lhc}. The SM Higgs decay branching ratios, including QCD 
corrections~\cite{zerwas}, have been obtained using the {\sc Hdecay} 
program~\cite{hdecay}. 
Results are summarised in Figure~\ref{fig:sum}. The $c$-quark mass
and the $\alpha_s$ uncertainties limit the predictivity of the $c \bar c$
and $g g$ channels to about $\pm 14\%$ and $\pm 7\%$ respectively.
On the contrary, the $b \bar b$, $\tau^+ \tau^-$ and $W W^*$ predictions can 
be obtained with accuracies comparable to, or better than, the 
experimental uncertainties.

For comparing these SM predictions to those in MSSM, a scan of the MSSM 
parameters phase space has been performed assuming:
\begin{center}
$2 < \tan \beta < 60$\\
$150~{\mathrm GeV}/c^2 < M_{A} < 1100~{\mathrm GeV}/c^2$\\
$500~{\mathrm GeV}/c^2 < M_{SUSY} < 1500~{\mathrm GeV}/c^2$\\
$-1000~{\mathrm GeV} < \mu < 1000~{\mathrm GeV}$\\
$0 < M^t_{LR}/M_{\tilde q} < \sqrt{6}$\\
$0.5 < M_{\tilde g}/M_{SUSY} < 1$.
\end{center} 
For each set of parameters the resulting $h^0$ mass has been computed using the
diagrammatic two-loop result~\cite{fhiggs1} implemented in the 
{\sc FeynHiggs} program~\cite{fhiggs2}. Solutions corresponding to 
$M_{h^0} = (120 \pm 2)$~GeV/$c^2$ have been selected and used to compute the 
$h^0$ decay branching ratios with the {\sc Hdecay} program, accounting for
squark loops, after forcing $M_{h^0}$ = 120~GeV/$c^2$. 
The pull quantity $\Delta (BR) = \frac{| BR^{MSSM} - BR^{SM}|}
{\sqrt{\sigma_{th}^2 + \sigma_{exp}^2}}$ has been computed for 
i) BR($h \rightarrow b \bar b$)/BR($h \rightarrow$ hadrons),
ii) BR($h \rightarrow c \bar c$)/BR($h \rightarrow$ hadrons),
iii) BR($h \rightarrow g \bar g$)/BR($h \rightarrow$ hadrons) and
iv) BR($h \rightarrow b \bar b$)/BR($h \rightarrow W W^*$). From the total 
$\chi^2$, the SM/MSSM discrimination has been defined as the portion of the 
$M_A - \tan \beta$ plane with more than 68\%, 90\% or 95\% of the MSSM 
solutions outside the SM 95\% confidence level region. 
This study has been repeated for four different scenarios: i) 500 fb$^{-1}$, 
theory systematics as above with CDR Vertex Tracker,
ii) 500 fb$^{-1}$, theory systematics as above with improved Vertex Tracker,
iii) 1000 fb$^{-1}$, theory systematics half of the above with CDR Vertex 
Tracker and iv) 1000 fb$^{-1}$, theory systematics half of the above with 
improved Vertex Tracker. SM/MSSM 90\% C.L. discrimination can be achieved up 
to $M_A \simeq$ 550~GeV/$c^2$ for the scenario i) improving to 
$M_A \simeq$ 730~GeV/$c^2$ assuming scenario iv). Since, with the experimental 
performances obtained for this analysis, the dominant source of uncertainties 
is due to the theory systematics, there is not a significant difference in the
excluded region between the two assumed performances for the Vertex Tracker.
The results are exemplified in Figure~\ref{fig:scan}. 
From this comparison it follows that the sensitivity to the SM/MSSM nature of 
a neutral Higgs boson is mainly limited by the theory uncertainties rather
than by the envisaged detector response and the experimental accuracies 
obtained in the present study.

\begin{figure}[ht!]
\begin{center}
\epsfig{file=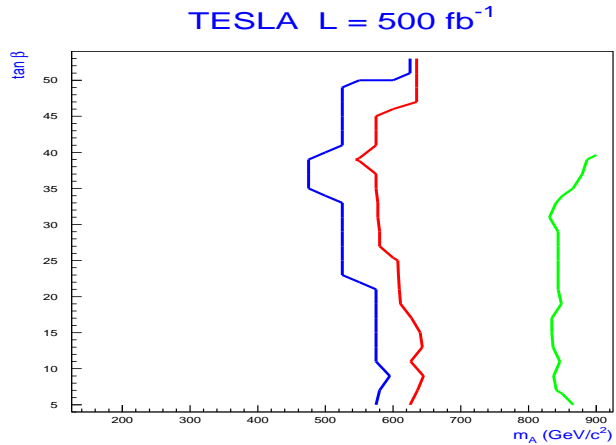,width=9.0cm,height=6.0cm}
\end{center}
\caption{The SM/MSSM discrimination region in the 
$M_A - \tan \beta$ plane for $M_H$ = 120~GeV/$c^2$ and 500 fb$^{-1}$. 
The lines define the regions with 68\% (light grey), 90\% (medium grey) and 
95\% (dark grey) of the MSSM solutions distinguishable at the 95\% confidence
level from the SM $H^0$ boson. \label{fig:scan}}
\end{figure}

It is also possible to use the accurate determination of the Higgs decay
branching ratios for an indirect estimate of the mass $M_{A^0}$ in the MSSM 
case. Sensitivity
to the $A^0$ mass arises from the MSSM corrections to the Higgs couplings
discussed above and it is of special interests for those masses above the
kinematical limit for direct $e^+e^- \rightarrow h^0 A^0$, $H^0 A^0$ 
production~\cite{ha}. 
The analysis has been performed assuming given sets of measured values for the 
BR($h \rightarrow c \bar c + g g$)/BR($h \rightarrow b \bar b$) and 
BR($h \rightarrow W W^*$)/BR($h \rightarrow b \bar b$) ratios. The $A^0$ mass 
has been varied together with the other MSSM parameters within the range 
compatible with the measured branching ratios allowing for their total 
uncertainty. The range of values of $M_A$ for the accepted MSSM solutions 
corresponded to an accuracy of 70~GeV/$c^2$ to 90~GeV/$c^2$ for the indirect
determination of $M_A$ in the mass range 300~GeV/$c^2 < M_A <$ 500~GeV/$c^2$.

\section{Conclusions}

The accuracy on the determination of the decay branching ratios for a neutral
Higgs boson with $M_{H^0}$ = 120 GeV/$c^2$ has been studied at a 
$\sqrt{s}$ = 350~GeV and 500~GeV $e^+ e^-$ linear collider. 
The measured branching ratios allow to determine the Higgs decay
width with good accuracy, to distinguish a MSSM $h^0$ boson from the SM
$H^0$ boson and to indirectly determine the $A^0$ mass up to 600-700~GeV/c$^2$.
Significant improvements in the knowledge of the $b$ and $c$ quark 
masses and good control of QCD corrections are necessary in order to 
preserve the predictive potential of the $c \bar c$ and $gg$ decays for 
$M_H <$ 120~GeV/c$^2$. For higher Higgs masses, the combination of the 
$b \bar b$ and $W W^*$ decay branching ratios provides the bulk of the 
information on the neutral Higgs nature and the $A^0$ mass.

\section*{Acknowledgements}

It is a pleasure to thank G.~Borisov, A.~Djouadi, E.~Gross, S.~Moretti, 
R.~Orava, M.~Peskin, F.~Richard, M.~Spira, D.~Treille and P.~Zerwas for 
suggestions and fruitful discussion. 
D.~Schulte has kindly provided the simulation of beamstrahlung
effect. I am also grateful to the organisers of the LCWS~99 Workshop for their 
invitation and the inspiring atmosphere during the conference.


\vspace{0.5cm}

\section*{References}
\begin{thebibliography}{99}

\bibitem{janot}
P. Janot, in {\it $e^+e^-$ collisions at 500~GeV: The physics potential},
DESY~92-123~A.

\bibitem{cdr}
{\it Conceptual Design of a 500 GeV $e^+e^-$ Linear Collider with Integrated 
X-ray Laser Facility} (ed. R. Brinkmann, G. Materlink, J. Rossbach and 
A. Wagner), DESY 1997-048.

\bibitem{vt}
M. Caccia {\it et al.}; T. Greenshaw {\it et al.}, in these Proceedings.

\bibitem{jetset}
T. Sjostrand, \Journal {\CPC}{82}{74}{1994}.

\bibitem{daniel}
D. Schulte, private communication.

\bibitem{simdet}
{\sc Simdet} program by H.J. Schreiber {\it et al.}

\bibitem{recoil}
W. Lohmann {\it et al.}, in these Proceedings.

\bibitem{hildreth}
M.D. Hildreth, T.L. Barklow and D.L. Burke, \Journal {\PRL}{49}{3441}{1994}.

\bibitem{hccgg}
I. Nakamura and K. Kawagoe, in Proc. of the {\it Workshop on 
Physics and Experiments with Linear Colliders}, vol. II, 
World Scientific, Singapore 1996.

\bibitem{desy123}
M. Battaglia and R. Vuopionpera, in {\it $e^+e^-$ Linear Colliders:
Physics and Detector Studies}, DESY~97-123~E.

\bibitem{taglep}
G. Borisov, \Journal {\NIMA}{417}{384}{1998},\\
M. Battaglia and D. Liko, {\it Identification of hadronic final states in $W$ 
decays by jet flavour tagging}, DELPHI~97-68~PHYS~614.

\bibitem{tagslc}
D. Jackson, \Journal {\NIMA}{388}{247}{1997}.

\bibitem{ww}
G. Borisov and F. Richard, {\it Precise measurement of Higgs decay rate into 
$W W^*$ at future $e^+e^-$ Linear Colliders}, LAL 99-26, hep-ph/9905413.

\bibitem{smssm}
H.E. Haber, in Proc. of the {\it  $4^{th}$ Int. Conf. on Physics beyond the 
Standard Model}, Lake Tahoe , CA, USA; World Scientific, Singapore, 1995;\\
J. Kamoshita, Y. Okada and M. Tanaka, in Proc. of the {\it Workshop on 
Physics and Experiments with Linear Colliders}, Morioka , Japan; vol. II, 
World Scientific, Singapore 1996;\\
J.F. Gunion, L. Poggioli and R. Van Kooten, in Proc. of the {\it 1996 DPF/DPB 
Summer Study on New Directions for High-energy Physics}, Snowmass, CO, USA;
APS, New York, 1997.
\bibitem{mb}
A. Hoang, {\it Phys. Rev.} {\bf D 59} (1998), 014039.

\bibitem{mc}
I. Bigi, M. Shifman and N. Uraltsev, {\it Aspects Of Heavy Quark Theory}, 
TPI-MINN-97-02-T (hep-ph/9703290).

\bibitem{top}
A. Hoang, in these Proceedings.

\bibitem{zerwas}
A. Djouadi, M. Spira and P. Zerwas, \Journal {\ZPC}{70}{427}{1996}.

\bibitem{hdecay}
A. Djouadi, J. Kalinowski and M. Spira, {\it HDECAY: A Program for Higgs Boson
Decays in the Standard Model and its Supersymmetric Extension}, DESY~97-079.

\bibitem{fhiggs1}
S. Heinemeyer, W. Hollik and G. Weiglein, \Journal {\PLB}{440}{296}{1998},
and \Journal {\PRD}{58}{091701}{1998}.

\bibitem{fhiggs2}
S. Heinemeyer, W. Hollik and G. Weiglein, {\it FeynHiggs: a program for the 
calculation of the masses of the neutral CP-even Higgs bosons in the MSSM}, 
CERN-TH/98-389.

\bibitem{ha}
A. Andreazza and C. Troncon, in {\it $e^+e^-$ Linear Colliders:
Physics and Detector Studies}, DESY~97-123~E.

\end{thebibliography}
\end{document}